\documentclass[aps,prl,twocolumn,showpacs,preprintnumbers,amsmath,amssymb]{revtex4}
\usepackage{graphicx}% Include figure files
\usepackage{dcolumn}% Align table columns on decimal point
\usepackage{mathrsfs}
\usepackage{bm}% bold math

\usepackage{amsfonts}
\usepackage{amssymb}
\usepackage{amsmath}
\usepackage{color}

\begin{document}

\preprint{}

\title{Modulation and Measurement of Time-Energy Entangled Photons}

\author{Chinmay Belthangady}
\email{chinmayb@stanford.edu}
\author{Shengwang Du}
\altaffiliation{Current address: Department of Physics, The Hong Kong University of Science and Technology, Clear Water Bay, Kowloon, Hong Kong, China.}
\author{Chih-Sung Chuu}
\author{G.Y. Yin}
\author{S.E. Harris}
\address{
Edward L. Ginzton Laboratory, Stanford University, Stanford,
California 94305, USA}
\date{\today}% It is always \today, today,

\begin{abstract} We describe a proof-of-principal experiment demonstrating a Fourier technique for measuring the shape of biphoton wavepackets.   The technique is based on the use of synchronously driven fast modulators and slow (integrating) detectors. 
\end{abstract}

\maketitle

The characterization of the joint quantum state of  time-energy entangled photons is a long-standing problem of interest to workers in quantum information processing\cite {Walmsley, Howell}. This interest is, in part, motivated by the quantum property where either dispersion \cite{Franson} or modulation \cite{Harris; nonlocal} that is experienced by a photon in one channel may be negated by the dispersion or modulation that is experienced by a second photon in a different channel.  By using appropriate spectral tools, the measurement of frequency domain correlation is readily accomplished. The measurement of temporal correlation is more difficult: If the temporal resolution of the photon detector is high enough, then a time-to-digital converter (TDC) may be used to directly measure it \cite {Shih; dispersive, Balic}. When the detectors do not have sufficient speed the measurement alternatives are limited. The technique of Hong-Ou-Mandel \cite{HOM} does not respond to even order dispersion \cite{Steinberg} and, though often used for this purpose, only measures the biphoton waveform when the waveform is transform limited. When sufficient count rate is available, sum frequency generation in a nonlinear crystal may be used as an ultra-fast correlator \cite{Silberberg; correlation, SensarnSum}. If the down-converting crystal is synchronously pumped, then femtosecond time-scale up-conversion may be used for temporal correlation \cite{Wong}.

\begin{figure}[htbp]
\begin{center}
\includegraphics*[width=2.5 truein]{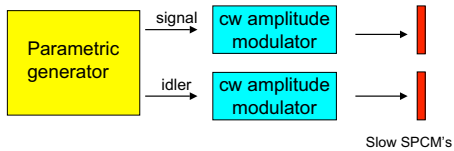}
\end{center}
\caption{Measurement of biphoton wave functions. Amplitude modulators in the signal and idler paths are driven synchronously. The coincidence count rate versus frequency is measured by slow detectors.}
\label{fig:schematic}
\end{figure}

This Letter describes a proof-of-principal experiment that demonstrates a new approach to the problem of measurement of biphoton wavefunctions.  Fig.~1 shows the essential idea.  Spontaneously generated signal and idler photons, or as in the experimental portion of this work, Stokes and anti-Stokes photons,
are incident on synchronously driven sinusoidal amplitude modulators.  The coincidence count rate between single photon counting modules (SPCMs) is measured as a function of the sinusoidal modulation frequency.  The SPCMs are slow as compared to the pulse width of the biphoton wave function, so that they integrate over the wavefunction.  With $\tau$ equal to the relative arrival time of the signal and idler photons, the inverse Fourier transform of the (frequency domain) measurement of coincidence counts versus frequency yields the Glauber  correlation function $G^{2}(\tau)$ and therefore the square of the absolute value of the biphoton wavefunction \cite{Harris; nonlocal}.  By definition, time to frequency Fourier transformation is the result of multiplication by a sinusoidal function and integration over time. Here, this operation is accomplished by modulation (multiplication) followed by slow detection (integration). 

In recent experiments, Kolchin et al. \cite{EOM} have demonstrated  electro-optic modulation of single photons by using one photon of a biphoton pair to set the time origin for the modulation of the second photon. Here, modulation of the biphoton wavefunction is attained without establishing a time origin for the modulator; i.e., the absolute common phase of the sinusoidal modulating signal is of no consequence.  Entangled signal and idler photons may be thought of as arriving at the modulators at the same, but random, time.  If the modulators are driven at different frequencies, or one channel is modulated and the other is not, then the Glauber correlation function is not changed by the modulators.  If the modulators are driven at the same frequency, then, as will be shown below, the  coincidence count rate is modulated. We note that some aspects of modulation of biphotons have been  considered by Tsang et al. \cite{Tsang}.

 \textsl{Theory}: We work in the Heisenberg picture with expectations taken against the vacuum state at the input of the parametric generator \cite{Gatti1}. We let $\hat{a}_{s,0}(t)$ and $\hat{a}_{i,0}(t)$ denote the signal and idler operators at the output of the parametric generator and before the amplitude modulators. With $t$ and $t+\tau$ denoting the arrival times of the signal and idler photons at their respective detectors, the unmodulated Glauber intensity correlation function is then $G^{(2)}_{0}(\tau)=\langle\hat{a}^{\dag}_{i,0}(t)\hat{a}^{\dag}_{s,0}(t+\tau)\hat{a}_{s,0}(t+\tau)\hat{a}_{i,0}(t)\rangle$.  When amplitude modulators $m_{1}(t)$ and $m_{2}(t)$ are inserted between the downconversion source and the detectors, the field operators are modulated so that $\hat{a}_{s}(t)=m_{1}(t)\hat{a}_{s,0}(t)$ and $\hat{a}_{i}(t)=m_{2}(t)\hat{a}_{i,0}(t)$. The modulated correlation function is then
\begin{equation}
G^{(2)}_{M}(t, t+\tau) = \left | m_{1}(t)\right |^{2}\left |m_{2}(t+\tau)\right |^{2}G^{(2)}_{0}(\tau)
\end{equation}

For a source with time independent statistics the unmodulated correlation function $ G^{(2)}_{0}(\tau)$ is a function of the difference of the  detection times of the two photons $\tau$ and does not depend explicitly on the arrival time $t$ of the first photon.  But as a result of the random time of arrival of the photon pair with regard to the phased modulators, the correlation function of the modulated biphoton does depend on both $t$ and $\tau$.  We average $G^{(2)}_{M}(t, t+\tau)$ over a period 
$\rm{T}$ of the modulating frequency to form the time averaged correlation function
\begin{equation}
\overline{G^{(2)}_{M}(\tau)}= \frac{1}{T}\int^{T}_{0} G^{(2)}_{M}(t, t+\tau)   dt.
\end{equation}

Combining Eq. (1) and Eq.(2), we obtain
\begin{eqnarray}
\label{m_defn}
&&\overline{G^{(2)}_{M}(\tau)}=\mathcal{M}(\tau) G^{(2)}_{0}(\tau) \nonumber \\
&&\mathcal{M}(\tau)= \frac{1}{T}\int^{T}_{0}\left |m_{1}(t)\right|^{2}\left |m_{2}(t+\tau)\right|^{2} dt.
\end{eqnarray}
where $\mathcal{M}(\tau)$  is the intensity correlation function of the modulators in the signal and idler channels. We thus have the result that the time-averaged Glauber correlation function of the modulated biphoton wavefunction is equal to the correlation function of the modulators multiplied by the unmodulated Glauber correlation function.

If one channel is modulated and the other is not, $\mathcal{M}(\tau)=1/2$, and the intensity correlation function is not modulated. If both channels are modulated by sinusoidal amplitude modulators with frequency $\omega$ and a common phase $\varphi$, that is by modulators 
$m_{1} (t)=m_{2} (t)=\cos(\omega t+\varphi)$, then irrespective of this phase, $\mathcal{M}(\tau)=1/4+1/8 \cos(2\omega \tau)$.

The essence of the Fourier technique is now apparent: If the detectors are slow in the sense that they integrate the coincidence counts for a time that is long as compared the the length of the biphoton wave packet , but short as compared to the inverse rate of biphoton generation, then the integrated coincidence count is 
$\int^{\infty}_{0}\overline{G^{(2)}_{M}(\tau)} d\tau=1/8 \int^{\infty}_{0}  \left [2+ \cos(2\omega \tau) \right] G^{(2)}_{0}(\tau)  d\tau$. We neglect the DC term, and normalize to obtain the Fourier cosine transform pair
\begin{eqnarray}
\label{Fourier}
&&F(2 \omega)= \sqrt{\frac{2}{\pi}}\int^{\infty}_{0} G^{(2)}_{0}(\tau) \cos(2 \omega \tau) d\tau\nonumber\\
&&G^{(2)}_{0}(\tau)=\sqrt{\frac{2}{\pi}}\int^{\infty}_{0}  F(2 \omega)  \cos(2 \omega \tau) d\omega
\end{eqnarray}
With the unmodulated biphoton wavefunction denoted by $\phi(\tau)$, and taking the rate of generated paired photons as R, then as shown by Rubin et al. \cite{Rubin94} and Kolchin \cite{ KolchinPRA},
$G^{(2)}_{0}(\tau)=|\phi(\tau)|^2+R^2$.  Assuming a bin width $\delta T$, and  photodetectors having a quantum efficiency of $\epsilon$, the coincidence count rate $R_c(\tau)$ is related to $G^{(2)}_{0}(\tau)$ by $R_c(\tau)=(\epsilon^2  \delta T  ) G^{(2)}_{0}(\tau)$.

\begin{figure}[htbp]
\begin{center}
\includegraphics*[width=2.5 truein]{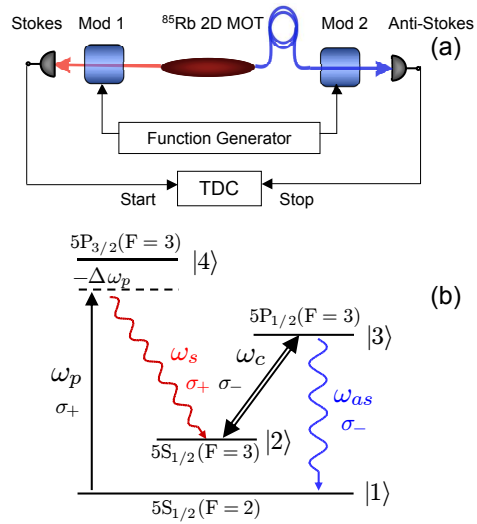}
\end{center}
\caption{(color online) (a) Schematic of experiment. Long biphotons are generated in the cold-atom Rb cell. Using an optical fiber, the anti-Stokes photon is delayed by 175 ns and the photons are modulated by synchronously driven sinusoidal modulators (b) Energy level diagram for paired photon generation in Rb.}
\label{fig:schematic}
\end{figure}

 \textsl{Experiment}: Our experiment makes use of long biphotons that are produced using the techniques of electromagnetically induced transparency  and slow light. The use of long biphotons allows us to compare the correlation function measured by our Fourier transform technique with a direct measurement using fast detectors and a TDC. The experimental configuration is shown in Fig. 2 (a). Paired photons are generated with cold Rb atoms using the method of Balic et al. \cite{Balic}. We apply strong counterpropagating pump and coupling lasers (not shown) to produce phase matched counter-propagating pairs of time-energy entangled Stokes and anti-Stokes photons. We use a $^{85}$Rb two-dimensional magneto-optic trap with an optical depth which can be varied between 10 to 60 \cite{subnatural} to generate biphotons with temporal lengths between 50 and 900 ns. The inset in Fig. 3(a) shows the biphoton wavefunction obtained at an optical depth of 35 as directly measured using a TDC.  Two features are of interest. First, the width of the wavefunction is determined by the slow group velocity of the anti-Stokes photon and varies linearly with the optical depth \cite{subnatural}. Second, the distinctive sharp feature at the leading edge is a Sommerfield-Brillouin precursor that ensures that the earliest signal reaches a detector at the speed of light in vacuum \cite{precursor}.

The generated Stokes and anti-Stokes  photons are transmitted through 10 GHz electro-optic amplitude modulators (Eospace Inc.) with a half-wave voltage, $V_{\pi}$ of 1.3V. To obtain a perfectly sinusoidal output, the modulators are biased at maximum transmission and the input voltage is varied linearly using a triangular waveform that varies between $-V_{\pi}$ and $+V_{\pi}$. This waveform is generated by a fast function generator (Tektronics AFG3252) with two output channels whose frequencies and phases can be varied independently. The modulated photons are then sent to SPCMs (Perkin Elmer SPCM-AQR-13), which are connected to the start and stop inputs of a TDC (Fast-Comtec TDC 7886S). Coincidence counts are binned into histograms and plotted as a function of the time difference between the detection of a Stokes and an anti-Stokes photon. 

\begin{figure}[t]
\begin{center}
\includegraphics*[width=3 truein]{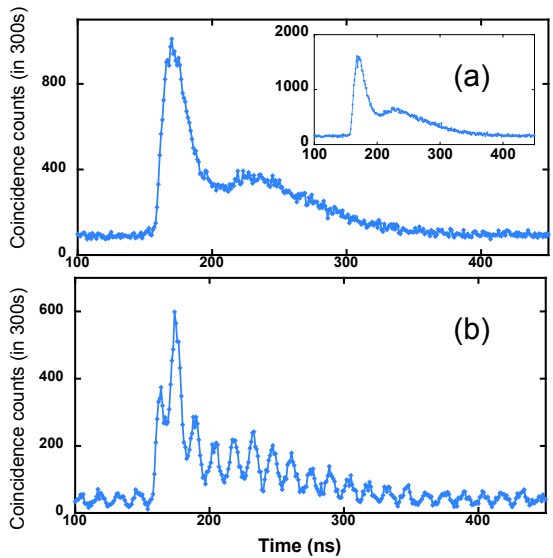}
\end{center}
\caption{(color online) Modulation of the biphoton correlation function. (a) $m_{1}(t)=1$ and $m_{2}(t)=\cos(\omega t+\varphi)$. The inset in (a) shows the correlation function with both modulators open. (b) $m_{1}(t)=m_{2}(t)=\cos(\omega t+\varphi)$. Here $\omega = 2\pi \times$ 35 $\times 10^{6}$.} 
\label{fig:modulation}
\end{figure}

In Fig. \ref{fig:modulation} (color online) we demonstrate the modulation of a biphoton wavefunction. The data are recoded by binning coincidence counts versus time into 1 ns bins. In Fig. \ref{fig:modulation}(a), the Stokes modulator is turned off and is biased at maximum transmission; and the anti-Stokes modulator is driven at  35 MHz.  In agreement with theory, the biphoton wavefunction is not modulated. In Fig.\ref{fig:modulation}(b), both modulators are modulated at 35 MHz with the same, but arbitrary, phase. The correlation function is now modulated at twice the applied frequency,  i.e. at 70 MHz. When driven by non-sinusoidal waveforms, the correlation function is modulated by the cross-correlation of the two modulating signals. In agreement with Eq. [3], we have verified that when the two modulating signals are square waves with the same frequency, the correlation function is modulated by a triangular function. 

\begin{figure}[t]
\begin{center}
\includegraphics*[width=3 truein]{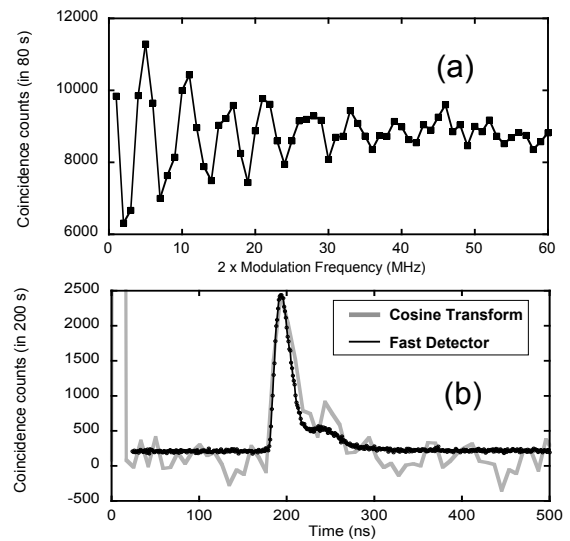}
\end{center}
\caption{ (color online) Fourier Transform measurement technique at an optical depth of 15. (a) Frequency domain data, (b) Fourier Cosine transform (gray), and temporal correlation (black). The term modulation frequency on the x-axis of (a) refers to the applied frequency. The observed modulation frequency [Eq.(4)] is a factor of 2 higher. }
\label{fig:gaussian}
\end{figure}

We next demonstrate the Fourier measurement technique.  As shown in Fig.  \ref{fig:schematic}(a), we use a 35m long polarization maintaining fiber to delay the anti-Stokes photon by 175 ns. The modulators are driven synchronously at frequencies between 0 MHz and 30 MHz. From the Fourier transform property that delay in the time domain corresponds to oscillation in the frequency domain, the plot of coincidence counts versus modulation frequency shows ripples with a frequency (5.7 MHz) equal to the inverse of the time delay. To simulate slow detectors we bin coincidences into 1$\mu$s bins, and since the temporal extent of the waveform is less than 1$\mu$s, the entire biphoton is contained within the first time bin. As a first example, we set the MOT optical depth to 15, the coupling laser Rabi frequency, $\Omega_{c}=3.47 \Gamma_{31}$ and the pump laser Rabi frequency, $\Omega_{p}=4.74 \Gamma_{31}$, where $\Gamma_{31}= (2 \pi) \times 6\times10^6 $ is the spontaneous decay rate out of state $|3\rangle$. For these parameters, the correlation function has an approximately Gaussian-like shape. In Fig. \ref{fig:gaussian}(a), we plot the number of coincidence counts in the first 1$\mu$s bin as a function of the modulation frequency.  For this plot, the experimental rate of paired photons is 325 s$^{-1}$ and for each point data is collected for 80s. In Fig. \ref{fig:gaussian}(b), we compare the biphoton wavefunction as  directly measured  with 1 ns bins (black line) to the Fourier cosine transform of the trace of Fig. \ref{fig:gaussian}(a) (gray line) vertically scaled for a one point fit. We find reasonable agreement between the two methods. We note that the sharp spike in the temporal trace near $t=0$ results from the DC component in the frequency domain trace. 

\begin{figure}[t]
\begin{center}
\includegraphics*[width=3 truein]{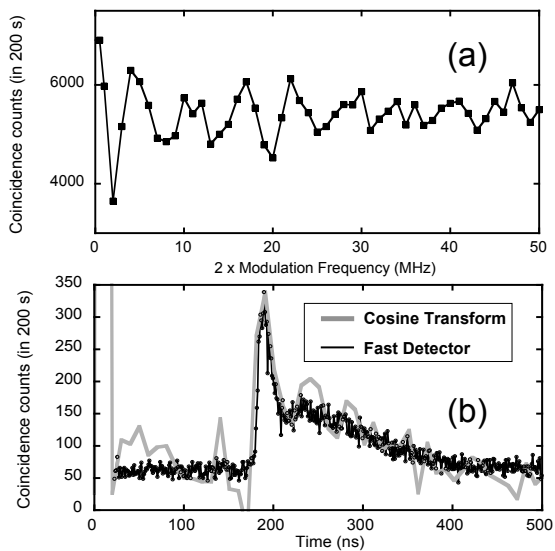}
\end{center}
\caption{(color online) Fourier Transform measurement technique at an optical depth of 35. (a) Frequency domain data, (b) Fourier Cosine transform (grey), and temporal correlation (black).}
\label{fig:groupdelay}
\end{figure}

We next increase the optical depth to 35 and set $\Omega_{c}=2.45 \Gamma_{31}$ and  $\Omega_{p}=3.21 \Gamma_{31}$.  The biphoton wavefunction is now characterized by a leading edge spike and an almost rectangular main body.  Fig.\ref{fig:groupdelay}(a) shows coincidence counts collected in the first 1$\mu$s bin plotted as a function of modulator frequency. Here, the experimental rate of paired counts is 80$s^{-1}$ and for each point, data is collected for 200s. The cosine transform of this trace, multiplied by a vertical scaling factor, is plotted against time in Fig. \ref{fig:groupdelay}(b) (gray line). For comparison, we show the correlation function where coincidences have been histogrammed into 1 ns bins.  The shapes measured by the two techniques are once again in reasonable agreement. 

In summary we have shown how biphotons may be modulated, and how this modulation may be used to measure the magnitude of the biphoton wavefunction. Though we have used long biphotons and low modulation frequencies, we believe that this technique should be extendable to short biphotons. A commercially available telecommunication modulator driven at a frequency of 60 GHz and therefore modulating at 120 GHz [Eq. (4)] will allow measurement of biphotons with a minimum length of about 8 picoseconds. This is about a factor of five faster than state-of-the-art commercial SPCMs. If the comparison is made on the basis of state-of-the-art polymer light modulators operating at 200 GHz \cite{polymer} and therefore modulating at 400 GHz, then the Fourier technique, will allow measurement down to about 2.5 ps. This is about eight times faster than the fastest reported superconducting detector \cite{superconducting}.  Looking further to the future, all-optical light modulators have been demonstrated at frequencies greater than 1 THz \cite{all optical}; thereby in principle allowing measurement of photons on femtosecond time scales. We note that an advantage of the  Fourier technique as compared to either sum frequency correlation or Hong-Ou-Mandel interference is that the signal and idler photons may be correlated at distant detectors and do not need to be brought together at a summing crystal or at a beam splitter. 

The authors acknowledge helpful discussions with S. Sensarn and P. Kolchin.  P. Kolchin also made substantial contributions to the apparatus used in this work. This work was supported by the U.S. Army Research Office, the U.S. Air Force Office of ScientiÞc Research, and the Defense Advanced Research Projects Agency.

\end{document}